\documentclass[sigconf, nonacm]{acmart}
\pdfoutput=1

\newcommand\vldbdoi{XX.XX/XXX.XX}

\newcommand\vldbavailabilityurl{URL_TO_YOUR_ARTIFACTS}
\newcommand\vldbpagestyle{plain}

\usepackage{caption}
\usepackage{amsthm}
\usepackage[inline]{enumitem}
\usepackage{pgfplots}
\usepackage{adjustbox}
\usepackage{layouts}
\pgfplotsset{compat=1.17}

\newtheorem{theorem}{Theorem}[section]
\newtheorem{lemma}[theorem]{Lemma}

\theoremstyle{definition}
\newtheorem{definition}{Definition}[section]

\newcommand{\Tau}{\mathcal{T}}

\begin{document}
\title{Efficient Approximate Search for Sets of Vectors}

%%
%% The "author" command and its associated commands are used to define the authors and their affiliations.
\author{Michael Leybovich}
\affiliation{%
  \institution{Technion}
  \city{Haifa}
  \state{Israel}
}
\email{smike87@cs.technion.ac.il}

\author{Oded Shmueli}
\affiliation{%
  \institution{Technion}
  \city{Haifa}
  \state{Israel}
}
\email{oshmu@cs.technion.ac.il}

\begin{abstract}
We consider a similarity measure between two sets $A$ and $B$ of vectors, that balances the average and maximum cosine distance between pairs of vectors, one from set $A$ and one from set $B$. As a motivation for this measure, we present lineage tracking in a database. To practically realize this measure, we need an approximate search algorithm that given a set of vectors $A$ and sets of vectors $B_1,...,B_n$, the algorithm quickly locates the set $B_i$ that maximizes the similarity measure. For the case where all sets are singleton sets, essentially each is a single vector, there are known efficient approximate search algorithms, e.g., approximated versions of tree search algorithms, locality-sensitive hashing (LSH), vector quantization (VQ) and proximity graph algorithms. In this work, we present approximate search algorithms for the general case. The underlying idea in these algorithms is encoding a set of vectors via a ``long'' single vector. The proposed approximate approach achieves significant performance gains over an optimized, exact search on vector sets.
\end{abstract}
\maketitle

% Front Matter
% ------------

%%% do not modify the following VLDB block %%
%%% VLDB block start %%%
\pagestyle{\vldbpagestyle}
% \begingroup\small\noindent\raggedright\textbf{PVLDB Reference Format:}\\
% \vldbauthors. \vldbtitle. PVLDB, \vldbvolume(\vldbissue): \vldbpages, \vldbyear.\\
% \href{https://doi.org/\vldbdoi}{doi:\vldbdoi}
% \endgroup
% \begingroup
% \renewcommand\thefootnote{}\footnote{\noindent
% This work is licensed under the Creative Commons BY-NC-ND 4.0 International License. Visit \url{https://creativecommons.org/licenses/by-nc-nd/4.0/} to view a copy of this license. For any use beyond those covered by this license, obtain permission by emailing \href{mailto:info@vldb.org}{info@vldb.org}. Copyright is held by the owner/author(s). Publication rights licensed to the VLDB Endowment. \\
% \raggedright Proceedings of the VLDB Endowment, Vol. \vldbvolume, No. \vldbissue\ %
% ISSN 2150-8097. \\
% \href{https://doi.org/\vldbdoi}{doi:\vldbdoi} \\
% }\addtocounter{footnote}{-1}\endgroup
%%% VLDB block end %%%

%%% do not modify the following VLDB block %%
%%% VLDB block start %%%
% \ifdefempty{\vldbavailabilityurl}{}{
% \vspace{.3cm}
% \begingroup\small\noindent\raggedright\textbf{PVLDB Artifact Availability:}\\
% The source code, data, and/or other artifacts have been made available at \url{\vldbavailabilityurl}.
% \endgroup
% }
%%% VLDB block end %%%

% Main Matter
% ------------
%
% To conform to Technion regulations, the main matter should begin
% with an introduction (including a survey of relevant past work):
%
\section{Introduction}
\label{sec:intro}

We consider a similarity measure between two sets $A$ and $B$ of vectors, that balances the average and maximum cosine distance between pairs of vectors, one from set $A$ and one from set $B$. Such a similarity measure is used in \cite{DBLP:conf/vldb/LeybovichS20}, in which we obtain a lineage via an embedding based method, such that each tuple/column is associated with a set of vectors. In \cite{DBLP:conf/vldb/LeybovichS20}, a lineage of a tuple is the set of tuples that explains its existence, either in a query result, or more generally, due to a sequence of database operations.
\par To employ the set-based similarity measure, we need to solve the following search problem. Given a set of vectors $A$ and sets of vectors $B_1,...,B_n$, where all vectors are of the same dimension $D$ (e.g., 200), efficiently locate the set $B_i$ that maximizes the similarity measure. A straightforward and inefficient implementation, would compare $A$ with each $B_i$. This deployment is infeasible for large databases where $n$ may be in the billions. We need a markedly more efficient search method. A similar problem was addressed by \cite{brecheisen2005efficient} which employs a very different similarity measure on two sets of vectors than the one we use. To mitigate the ``expensive'' operation of computing distance between set $A$ and each of the $B_i$ sets, they devised a ``cheap'' filter test that eliminates non-winners. However, they will still apply, at least the filter test, to all the $B_i$ sets.
\par For the case where all sets are singleton sets, essentially each is a single vector, there are known efficient approximate vector search algorithms to efficiently locate similar singleton sets of vectors. The most popular solutions are based on approximated versions of tree algorithms \cite{houle2014rank, muja2014scalable}, LSH \cite{lsh, DBLP:journals/corr/AndoniILRS15}, VQ \cite{jegou2010product, guo2016quantization, guo2020accelerating} and proximity graph algorithms \cite{arya1993approximate, malkov2018efficient}. To practically employ this measure, we aim to adapt such an approximate search algorithm to the current setting of set-based similarity.  
\par In the application of \cite{DBLP:conf/vldb/LeybovichS20}, the database tuples are each associated with a set of ``lineage vectors''. Given a ``target tuple'' which is associated with a set of ``lineage vectors'' $A$, whose ``existence'' is to be explained, such an efficient search algorithm would enable us to locate tuples, say $t_1,...,t_m$, for $i \in [1, m]$, where each tuple $t_i$ is associated with a set of lineage vectors $S_i$, such that $S_1,...,S_m$ are the closest, via the set-based similarity measure, to set $A$ amongst all the tuple-associated sets of vectors in the database.
\par In this work, we present approximate search algorithms for the general case of set-based similarity. The underlying idea in these algorithms is encoding a set of vectors via a ``long'' single vector. Once we represent a set via single long vectors, we can employ known approximate search techniques for vectors. Thus, we reduce the search problem on sets to one on single vectors.
\par We proceed in two steps. First, we outline a search method assuming that all sets of vectors have the same cardinality, say $N$, e.g., each such set has 3 vectors. Then, we generalize to the more complex case in which sets may have differing cardinalities. The construction is based on the principle of ``assumed winners''.
\par We prove formally the correctness of the proposed long vectors based exact search method; we conduct an empirical analysis for corroboration. Then, we present experimental results using an \textit{approximate} long vectors based search method, which significantly outperforms an optimized exact brute-force search on vector sets, while maintaining a very high recall score. This renders our method scalable to massive databases of sets of vectors.

\subsection*{Paper Organization}
% The rest of this paper is organized as follows.
Previous work on lineage via embeddings is briefly surveyed in Section \ref{chap:lineage_via_embeddings}, motivating the need for an efficient search for vector sets.
In Section \ref{sec:constant_cardinality_vector_search} we present a basic method under the assumption that all the sets of vectors are of a constant cardinality $N$.
Then, in Section \ref{sec:varying_cardinalities_vector_search} we introduce a refined version that supports sets of vectors of varying cardinalities, such that each set of vectors $V$, which is associated with a tuple or a tuple column in the database (DB), has a cardinality $|V| \in [1, max\_vectors\_num]$ ($max\_vectors\_num$ is a hyperparameter of our system).
In Section \ref{sec:correctness-proof}, we prove the correctness of the method described in Section \ref{sec:constant_cardinality_vector_search}.
In Section \ref{sec:experiments} we show experimental results that render our method practical for massive databases of sets of vectors.
% In Section \ref{sec:experiments} we present experimental results in a practical setting setting setting setting setting setting setting setting.
We conclude in Section \ref{chap:conclusions}.
\section{Motivation: Lineage via Embeddings}
\label{chap:lineage_via_embeddings}

\subsection{Outline}
% Our proposal - encode provenance information using word embeddings
\par In \cite{DBLP:conf/vldb/LeybovichS20} we devise a novel approach to lineage tracking, which is based on Machine Learning (ML) and Natural Language Processing (NLP) techniques. The main idea is summarizing, and thereby approximating, the lineage of every tuple or a tuple column via a small set of up to $max\_vectors\_num$ (a hyperparameter) constant-size vectors.
\par In NLP, \textit{word vectors} are vectors, whose elements are real numbers, which represent the ``semantic ties'' of a word in a text \cite{DBLP:journals/corr/abs-1301-3781, DBLP:conf/nips/MikolovSCCD13}.
Word vectors support similarity queries by grouping closely ``semantically'' similar words in a low-dimensional space.
Our sets of vectors encode lineage, and provide word vectors like ``grouping'' properties \cite{DBLP:conf/naacl/MikolovYZ13}. That is, given a set-set similarity metric, we support lineage querying by grouping closely the vector sets of related (lineage wise) tuples/columns.

% Explain the motivation and process of approx provenance querying
\subsection{Similarity calculation}\label{sec:similarity_calculation} Given two word vectors, the similarity score is usually the cosine distance between the vectors. In our system, we want to calculate the similarity between the lineage representations of two tuples/columns. In both cases the tuple, or column, is associated with a \textit{set} of lineage vectors. That is, we need to calculate the similarity between \textit{two sets of vectors}. 
We note that \cite{DBLP:journals/corr/BordawekarS16,DBLP:journals/corr/abs-1712-07199,DBLP:conf/sigmod/BordawekarS17}
have also used various similar methods for measuring similarity between two sets of vectors. The ``logic'' behind the following formula is balancing between the ``best pair'' of vectors (in terms of similarity) in the two sets and the similarity between their average vectors:\\
\begin{equation*}
    \operatorname{sim}(A, B) = \frac{w_{max} \cdot max(ps) + w_{avg} \cdot avg(ps)}
    {w_{max} + w_{avg}}\\
\end{equation*}

where $A$ and $B$ are sets of vectors, $ps$ is the set of pair-wise similarities between a pair of vectors, one taken from set $A$ and one taken from set $B$. $w_{max}$ and $w_{avg}$ are (user-specified) hyperparameters. $max$ and $avg$ are functions that return the maximum and average values of a collection of numbers, respectively. This logic holds for both tuple-based vectors and column-based vectors (i.e., vectors kept for each column separately).

\subsection{Lineage querying} Given a tuple and its lineage embedding vectors, we can calculate the pair-wise similarity against every other tuple in the DB (or a subset, e.g., in a specific table) and return the top $K$ (a parameter) most lineage-similar tuples (these resemble a subset of the lineage \cite{Cui:2000:TLV:357775.357777}). 
There are many algorithms for approximate vector search, e.g., based on LSH \cite{lsh}. Approximate vector search is a very active area of research and we can utilize known algorithms (see, e.g., \cite{sugawara-etal-2016-approximately}); however, these algorithms are not directly applicable to our method, since we operate on \textit{sets of vectors} instead of single vectors. In this paper we provide a practical reduction from the case of set-set
similarity to the case of vector-vector similarity. Therefore, as vector-vector similarity can be handled efficiently using known methods, we can
efficiently deploy a system using the set-set similarity measure
outlined above.\\

\section{Constant Cardinality of Sets of Vectors}\label{sec:constant_cardinality_vector_search}
Here, we assume that all sets of vectors are of a constant cardinality $N$. The idea is transforming each set of vectors into a single long vector, so that the \textit{dot product} of two such long vectors computes all pair-wise interactions, and is equivalent to the similarity calculation (between sets of vectors) we devised in section \ref{sec:similarity_calculation}. Note that all vectors are of the same constant dimension $D$ (a hyperparameter of our system).

\subsection{Long Vectors Construction}\label{sec:long_vectors_construction_constant}
\begin{enumerate}
    \item Let $A$ be a target set of vectors of cardinality $N$, for which we want to find the closest candidate set of vectors.
    \item Let $V$ be a candidate set of vectors for $A$ of cardinality $N$. Construct a \textit{long candidate vector} $\vec{L}_V \in \mathbb{R}^{|A| \times |V| \times D}$ from $V$ by concatenating $|A|$ copies of each (normalized) $\vec{v}_i \in V$, in order $\vec{v}_1, ..., \vec{v}_{|V|}$: \\
    \begin{equation*}
        \vec{L}_V = \begin{pmatrix}
                    \frac{\vec{v}_{1, 1}}{\lVert \vec{v}_1 \rVert}\\ 
                    \vdots\\ 
                    \frac{\vec{v}_{1, |A|}}{\lVert \vec{v}_1 \rVert}\\ 
                   \vdots\\ 
                   \frac{\vec{v}_{|V|, 1}}{\lVert \vec{v}_{|V|} \rVert}\\ 
                   \vdots\\ 
                   \frac{\vec{v}_{|V|, |A|}}{\lVert \vec{v}_{|V|} \rVert}\\
                   \end{pmatrix}
                  \underset{\scriptscriptstyle |A| = |V| = N}{=} 
                  \begin{pmatrix}
                    \frac{\vec{v}_{1, 1}}{\lVert \vec{v}_1 \rVert}\\ 
                    \vdots\\ 
                    \frac{\vec{v}_{1, N}}{\lVert \vec{v}_1 \rVert}\\ 
                   \vdots\\ 
                   \frac{\vec{v}_{N, 1}}{\lVert \vec{v}_N \rVert}\\ 
                   \vdots\\ 
                   \frac{\vec{v}_{N, N}}{\lVert \vec{v}_N \rVert}\\
                   \end{pmatrix}
                   \end{equation*}
        such that $\vec{v}_{i, j}$ is the $j^{th}$ copy of $\vec{v}_i$.
    \item Now, we build a long vector $\vec{L}_{A} \in \mathbb{R}^{|A| \times |V| \times D}$ by concatenating all (normalized) $\vec{a}_i \in A$, in order $\vec{a}_1, ..., \vec{a}_{|A|}$, and duplicating the result $|V|$ times: \\
    \begin{equation*}
        \vec{L}_{A} = \begin{pmatrix}
                    \frac{\vec{a}_{1, 1}}{\lVert \vec{a}_1 \rVert}\\ 
                    \vdots\\ 
                    \frac{\vec{a}_{|A|, 1}}{\lVert \vec{a}_{|A|} \rVert}\\ 
                   \vdots\\ 
                   \frac{\vec{a}_{1, |V|}}{\lVert \vec{a}_{1} \rVert}\\ 
                   \vdots\\ 
                   \frac{\vec{a}_{|A|, |V|}}{\lVert \vec{a}_{|A|} \rVert}\\
                   \end{pmatrix}
                  \underset{\scriptscriptstyle |A| = |V| = N}{=} 
                  \begin{pmatrix}
                     \frac{\vec{a}_{1, 1}}{\lVert \vec{a}_1 \rVert}\\ 
                    \vdots\\ 
                    \frac{\vec{a}_{N, 1}}{\lVert \vec{a}_{N} \rVert}\\ 
                   \vdots\\ 
                   \frac{\vec{a}_{1, N}}{\lVert \vec{a}_{1} \rVert}\\ 
                   \vdots\\ 
                   \frac{\vec{a}_{N, N}}{\lVert \vec{a}_{N} \rVert}\\
                   \end{pmatrix}
                   \end{equation*}
       such that $\vec{a}_{i, j}$ is the $j^{th}$ copy of $\vec{a}_i$.
    \item $\vec{L}_{A}$ is used to capture the \textit{average} of all pair-wise \textit{cosine similarities} with vectors from a candidate set $V$ (note that $a_{i, j} = a_i$ and $v_{j, i} = v_j$):\\
    \begin{equation*}
        \vec{L}_{A} \cdot \vec{L}_V = \sum_{i = 1}^{|A|}{\sum_{j = 1}^{|V|} \frac{\vec{a}_{i, j}}{\lVert \vec{a}_i \rVert} \cdot \frac{\vec{v}_{j, i}}{\lVert \vec{v}_j \rVert}} = sum(ps) =
    \end{equation*}
    \begin{equation*}
        = avg(ps) \times |A| \times |V| \underset{\scriptscriptstyle |A| = |V| = N}{=} avg(ps) \times N^2
    \end{equation*}
    where $ps$ is the multi-set of pair-wise similarities between a pair of vectors, one taken from set $V$ and one taken from set $A$.
    \item To capture the maximum of the pair-wise similarities (denoted $max(ps)$) we build $|A| \times |V| = N^2$ long ``selector'' vectors $\vec{\sigma}_{1, 1}, ..., \vec{\sigma}_{|A|, 1}, ..., \vec{\sigma}_{1, |V|}, ..., \vec{\sigma}_{|A|, |V|}$, each $\vec{\sigma}_{i, j} \in \mathbb{R}^{|A| \times |V| \times D}$ ``assumes'' which of the $|A| \times |V| = N^2$ pair-wise interactions is maximal:\\
    \begin{equation*}
        \vec{\sigma}_{i, j} = \begin{pmatrix}
                    \vec{0}_{1, 1}\\ 
                    \vdots\\ 
                    \vec{0}_{|A|, 1}\\ 
                   \vdots\\
                   \vec{1}_{i, j}\\ 
                   \vdots\\
                   \vec{0}_{1, |V|}\\ 
                   \vdots\\ 
                   \vec{0}_{|A|, |V|}\\ 
                   \end{pmatrix}
    \end{equation*}
    i.e., $\vec{\sigma}_{i, j}$ is a concatenation of $O_s = (j - 1) \times |A| + (i - 1)$ $\vec{0}$ vectors, followed by one $\vec{1}$ vector, and ending with $|A| \times |V| - (O_s + 1)$ $\vec{0}$ vectors,
    where $\vec{0} \in \mathbb{R}^D$ is the ``all zeros'' vector and $\vec{1} \in \mathbb{R}^D$ is the ``all ones'' vector. 
    $\vec{\sigma}_{i, j}$ ``assumes'' the maximum occurs in the cosine similarity product between $\vec{a}_i$ and $\vec{v}_j$.
    Consequently, we get:\\
    \begin{equation*}
        \vec{\sigma}_{i, j} \odot \vec{L}_{A} = \begin{pmatrix}
                    \vec{0}_{1, 1}\\ 
                    \vdots\\ 
                    \vec{0}_{|A|, 1}\\ 
                   \vdots\\
                   \frac{\vec{a}_{i, j}}{\lVert \vec{a}_i \rVert}\\ 
                   \vdots\\
                   \vec{0}_{1, |V|}\\ 
                   \vdots\\ 
                   \vec{0}_{|A|, |V|}\\ 
                   \end{pmatrix}
    \end{equation*}
    such that $\vec{a}_{i, j}$ is the $j^{th}$ copy of $\vec{a}_i$ and $\odot$ is the Hadamard (i.e., element-wise) product.
    This results in:
    \begin{equation*}
        (\vec{\sigma}_{i, j} \odot \vec{L}_{A}) \cdot \vec{L}_V = \frac{\vec{a}_{i, j}}{\lVert \vec{a}_i \rVert} \cdot \frac{\vec{v}_{j, i}}{\lVert \vec{v}_j \rVert} = \frac{\vec{a}_{i}}{\lVert \vec{a}_i \rVert} \cdot \frac{\vec{v}_{j}}{\lVert \vec{v}_j \rVert}
    \end{equation*}
    
    \item Next, we construct $|A| \times |V| = N^2$ \textit{long target vectors} $\vec{\tau}_{1, 1}, ..., $ $\vec{\tau}_{|A|, 1}, ..., \vec{\tau}_{1, |V|}, ..., \vec{\tau}_{|A|, |V|}$ ($\vec{\tau}_{i, j} \in \mathbb{R}^{|A| \times |V| \times D}$):
    \begin{equation*}
        \vec{\tau}_{i, j} =
        \frac{1}{w_{max} + w_{avg}} \cdot \left (
        w_{max} \cdot (\vec{\sigma}_{i, j} \odot \vec{L}_{A}) 
        + \frac{w_{avg}}{|A| \times |V|} \cdot \vec{L}_{A}
        \right )
    \end{equation*}
    where $w_{max}$ and $w_{avg}$ are (user-specified) hyperparameters, as specified in section \ref{sec:similarity_calculation}.
    \item Each long target vector $\vec{\tau}_{i, j}$ computes the desired similarity calculation via a dot product with a long candidate vector $\vec{L}_V$, under the assumption that $\vec{a}_i$ and $\vec{v}_j$ have the maximal pair-wise similarity:
    \begin{equation*}
         \vec{\tau}_{i, j} \cdot \vec{L}_V = \frac{w_{max} \cdot (\vec{\sigma}_{i, j} \odot \vec{L}_{A}) \cdot \vec{L}_V + \frac{w_{avg}}{|A| \times |V|} \cdot \vec{L}_{A} \cdot \vec{L}_V}{w_{max} + w_{avg}} 
        = 
    \end{equation*}
    \begin{equation*}
        = 
        \frac{w_{max} \cdot \frac{\vec{a}_{i, j}}{\lVert \vec{a}_i \rVert} \cdot \frac{\vec{v}_{j, i}}{\lVert \vec{v}_j \rVert} + w_{avg} \cdot avg(ps)}{w_{max} + w_{avg}}
        =
    \end{equation*}
    \begin{equation*}
        = 
        \frac{w_{max} \cdot \frac{\vec{a}_{i}}{\lVert \vec{a}_i \rVert} \cdot \frac{\vec{v}_{j}}{\lVert \vec{v}_j \rVert} + w_{avg} \cdot avg(ps)}{w_{max} + w_{avg}}
    \end{equation*}
    \label{calc:tau-dot-product-constant}
\end{enumerate}

\subsection{Search Method}\label{sec:search_method_constant}
\begin{enumerate}
    \item Insert all the long \textit{candidate} vectors (of a constant dimension $N^2 \times D$) into a vector search structure $S$. Note that known approximate vector search techniques, such as LSH \cite{lsh, DBLP:journals/corr/AndoniILRS15} (or any other technique, as mentioned in section \ref{sec:intro}), can be utilized here for efficiency. Search structure $S$ is search technique-specific.
    \item Recall that for a candidate set of vectors $V$, the long candidate vector is $\vec{L}_V$.
    \item Now, suppose you are given a target set of vectors $A$, of cardinality $N$. Construct $N^2$ long target vectors $\vec{\tau}_{1, 1}, ..., \vec{\tau}_{N, N}$. 
    \item We look \textit{separately} for the closest (dot-product wise) long candidate vector, denoted $\vec{L}_{{i, j}}$, in $S$, to each $\vec{\tau}_{i, j}$, respectively.\label{search:closest-dot-products-search}
    \item We compute the similarity scores $\vec{\tau}_{1, 1} \cdot \vec{L}_{{1, 1}}, ..., \vec{\tau}_{N, 1} \cdot \vec{L}_{{N, 1}}, ...,$ $\vec{\tau}_{1, N} \cdot \vec{L}_{{1, N}}, ..., \vec{\tau}_{N, N} \cdot \vec{L}_{{N, N}}$. The one yielding the highest score, say $\vec{L}_{{{\Tilde{i}, {\Tilde{j}}}}}$, identifies the desired candidate set of vectors $V_{{\Tilde{i}, {\Tilde{j}}}}$, according to our set-oriented similarity formula. If $u > 0$ closest sets are desired, the $u$ sets yielding the $u$ highest scores are output.
    \label{search:closest-dot-products-output}
\end{enumerate}

\section{Sets of Vectors of Varying Cardinalities}\label{sec:varying_cardinalities_vector_search}
In section \ref{sec:constant_cardinality_vector_search}, we presented the construction of a long candidate vector $\vec{L}_V$ from a candidate set of vectors $V$, and a collection of long target vectors $\vec{\tau}_{i, j}$ (where $i \in [1, |A|=N]$ and $j \in [1, |V| = N]$|) from a target set of vectors $A$, under the assumption that all sets of vectors are of a constant cardinality $N$. It is evident that these constructions are tightly coupled with the cardinalities of the target and candidate sets of vectors, namely $|A|$ and $|V|$.
Hence, we introduce a refined version that supports sets of vectors of varying cardinalities, such that each set of vectors
$V$ has a cardinality $|V| \in [1, max\_vectors\_num]$ ($max\_vectors\_num$ is hereafter denoted as $M$, for brevity). 
Recall that all vectors are of the same constant dimension $D$.
\par The general idea is pre-computing $M$ long candidate vectors $\vec{L}_V^1, ..., \vec{L}_V^M$ instead of a single $\vec{L}_V$, and a collection of long target vectors $\vec{\tau}_{i, j}^k$, for each $k \in [1, M]$ (where $i \in [1, |A|]$ and $j \in [1, k]$|). Each $\vec{L}_V^n \in \mathbb{R}^{n \times |V| \times D}$, where $n \in [1, M]$, ``assumes'' in its construction that the cardinality of a target set of vectors is $n$. Each $\vec{\tau}_{i, j}^k \in \mathbb{R}^{|A| \times k \times D}$, where $k \in [1, M]$, ``assumes'' in its construction that the cardinality of a candidate set of vectors is $k$. Consequently, instead of a \textit{single} form of long candidate and target vectors (of dimension $N^2 \times D$, as is the case in section \ref{sec:constant_cardinality_vector_search}), we potentially get $M \times M$ different such forms (in terms of dimension and $(n, k)$ construction parameters), such that each long vector of dimension $n \times k \times D$, where $n,k \in [1, M]$, is associated with a separate search structure $S_{1,1}, ..., S_{M,M}$, depending on its $(n, k)$ form (there are additional options, e.g., combining such search structures of equal dimensions, these will not be discussed further).

\subsection{Long Vectors Construction}
\begin{enumerate}

    \item Let $V$ be a candidate set of vectors of cardinality $|V| \in [1, M]$. Construct $M$ long candidate vectors $\vec{L}_V^1, ..., \vec{L}_V^M$, such that each $\vec{L}_V^n \in \mathbb{R}^{n \times |V| \times D}$, where $n \in [1, M]$, is constructed from $V$ by concatenating $n$ copies of each (normalized) $\vec{v}_i \in V$, in order $\vec{v}_1, ..., \vec{v}_{|V|}$: \\
    \begin{equation*}
        \vec{L}_V^n = \begin{pmatrix}
                    \frac{\vec{v}_{1, 1}}{\lVert \vec{v}_1 \rVert}\\ 
                    \vdots\\ 
                    \frac{\vec{v}_{1, n}}{\lVert \vec{v}_1 \rVert}\\ 
                   \vdots\\ 
                   \frac{\vec{v}_{|V|, 1}}{\lVert \vec{v}_{|V|} \rVert}\\ 
                   \vdots\\ 
                   \frac{\vec{v}_{|V|, n}}{\lVert \vec{v}_{|V|} \rVert}\\
                   \end{pmatrix}
                   \end{equation*}
        such that $\vec{v}_{i, j}$ is the $j^{th}$ copy of $\vec{v}_i$.
        $\vec{L}_V^n$ ``assumes'' the cardinality of the target set of vectors is $n$.
        
    \item Let $A$ be a target set of vectors of cardinality $|A| \in [1, M]$.

    \item Now, we build $M$ long vectors $\vec{L}_{A}^1, ..., \vec{L}_{A}^M$, such that $\vec{L}_{A}^k \in \mathbb{R}^{|A| \times k \times D}$, where $k \in [1, M]$, is built by concatenating all (normalized) $\vec{a}_i \in A$, in order $\vec{a}_1, ..., \vec{a}_{|A|}$, and duplicating the result $k$ times: \\
        \begin{equation*}
            \vec{L}_{A}^k = \begin{pmatrix}
                    \frac{\vec{a}_{1, 1}}{\lVert \vec{a}_1 \rVert}\\ 
                    \vdots\\ 
                    \frac{\vec{a}_{|A|, 1}}{\lVert \vec{a}_{|A|} \rVert}\\ 
                   \vdots\\ 
                   \frac{\vec{a}_{1, k}}{\lVert \vec{a}_{1} \rVert}\\ 
                   \vdots\\ 
                   \frac{\vec{a}_{|A|, k}}{\lVert \vec{a}_{|A|} \rVert}\\
                   \end{pmatrix}
       \end{equation*}
       such that $\vec{a}_{i, j}$ is the $j^{th}$ copy of $\vec{a}_i$.
       $\vec{L}_{A}^k$ ``assumes'' the cardinality of the candidate set of vectors is $k$.
    \item Let $V$ be a candidate set of vectors for $A$ of cardinality $|V| \in [1, M]$.
   Long vector $\vec{L}_{A}^{k=|V|} \in \mathbb{R}^{|A| \times |V| \times D}$ is used to capture the \textit{average} of all pair-wise \textit{cosine similarities} with vectors from $V$ ($\vec{L}_V^{n=|A|} \in \mathbb{R}^{|A| \times |V| \times D}$):\\
    \begin{equation*}
        \vec{L}_{A}^{k=|V|} \cdot \vec{L}_V^{n=|A|} =
        \vec{L}_{A}^{|V|} \cdot \vec{L}_V^{|A|} = \sum_{i = 1}^{|A|}{\sum_{j = 1}^{|V|} \frac{\vec{a}_{i, j}}{\lVert \vec{a}_i \rVert} \cdot \frac{\vec{v}_{j, i}}{\lVert \vec{v}_j \rVert}} =
    \end{equation*}
    \begin{equation*}
        =
        sum(ps) = avg(ps) \times |A| \times |V|
    \end{equation*}
    where $ps$ is the multi-set of pair-wise similarities between a pair of vectors, one taken from set $V$ and one taken from set $A$.
    
    \item To capture the maximum of the pair-wise similarities (denoted $max(ps)$) we build $|A| \times k$ long ``selector'' vectors $\vec{\sigma}_{1, 1}^k, ...,$ $\vec{\sigma}_{|A|, 1}^k, ..., \vec{\sigma}_{1, k}^k, ..., \vec{\sigma}_{|A|, k}^k$, for each $k \in [1, M]$. 
    That is, a total of $\sum_{k=1}^M |A| \times k = |A| \times \sum_{k=1}^M k = |A| \times \frac{M(1 + M)}{2}$ long ``selector'' vectors.
    Each $\vec{\sigma}_{i, j}^k \in \mathbb{R}^{|A| \times k \times D}$ ``assumes'' which of the $|A| \times k$ pair-wise interactions is maximal:\\
    \begin{equation*}
        \vec{\sigma}_{i, j}^k = \begin{pmatrix}
                    \vec{0}_{1, 1}\\ 
                    \vdots\\ 
                    \vec{0}_{|A|, 1}\\ 
                   \vdots\\
                   \vec{1}_{i, j}\\ 
                   \vdots\\
                   \vec{0}_{1, k}\\ 
                   \vdots\\ 
                   \vec{0}_{|A|, k}\\ 
                   \end{pmatrix}
    \end{equation*}
    i.e., $\vec{\sigma}_{i, j}^k$ is a concatenation of $O_s = (j - 1) \times |A| + (i - 1)$ $\vec{0}$ vectors, followed by one $\vec{1}$ vector, and ending with $|A| \times k - (O_s + 1)$ $\vec{0}$ vectors,
    where $\vec{0} \in \mathbb{R}^D$ is the ``all zeros'' vector and $\vec{1} \in \mathbb{R}^D$ is the ``all ones'' vector. 
    $\vec{\sigma}_{i, j}^k$ ``assumes'' the maximum occurs in the cosine similarity product between $\vec{a}_i$ and $\vec{v}_j$. Also, $\vec{\sigma}_{i, j}^k$ ``assumes'' the cardinality of the set of lineage vectors of the candidate tuple is $k$.
    Consequently, we get:\\
    \begin{equation*}
        \vec{\sigma}_{i, j}^k \odot \vec{L}_{A}^k = \begin{pmatrix}
                    \vec{0}_{1, 1}\\ 
                    \vdots\\ 
                    \vec{0}_{|A|, 1}\\ 
                   \vdots\\
                   \frac{\vec{a}_{i, j}}{\lVert \vec{a}_i \rVert}\\ 
                   \vdots\\
                   \vec{0}_{1, k}\\ 
                   \vdots\\ 
                   \vec{0}_{|A|, k}\\ 
                   \end{pmatrix}
    \end{equation*}
    such that $\vec{a}_{i, j}$ is the $j^{th}$ copy of $\vec{a}_i$ and $\odot$ is the Hadamard (i.e., element-wise) product.
    This results in:
    \begin{equation*}
        (\vec{\sigma}_{i, j}^{k=|V|} \odot \vec{L}_{A}^{k=|V|}) \cdot \vec{L}_V^{n=|A|} =
        (\vec{\sigma}_{i, j}^{|V|} \odot \vec{L}_{A}^{|V|}) \cdot \vec{L}_V^{|A|} =
    \end{equation*}
    \begin{equation*}
        = \frac{\vec{a}_{i, j}}{\lVert \vec{a}_i \rVert} \cdot \frac{\vec{v}_{j, i}}{\lVert \vec{v}_j \rVert} = \frac{\vec{a}_{i}}{\lVert \vec{a}_i \rVert} \cdot \frac{\vec{v}_{j}}{\lVert \vec{v}_j \rVert}
    \end{equation*}
    
    \item Next, we construct $|A| \times k$ long target vectors $\vec{\tau}_{1, 1}^k, ..., \vec{\tau}_{|A|, 1}^k, ..., $ $\vec{\tau}_{1, k}^k, ..., \vec{\tau}_{|A|, k}^k$ for each $k \in [1, M]$. That is, a total of $\sum_{k=1}^M |A| \times k = |A| \times \sum_{k=1}^M k = |A| \times \frac{M(1 + M)}{2}$ long target vectors $\vec{\tau}_{i, j}^k \in \mathbb{R}^{|A| \times k \times D}$:
    \begin{equation*}
        \vec{\tau}_{i, j}^k =
        \frac{1}{w_{max} + w_{avg}} \cdot \left (
        w_{max} \cdot (\vec{\sigma}_{i, j}^k \odot \vec{L}_{A}^k) 
        + \frac{w_{avg}}{|A| \times k} \cdot \vec{L}_{A}^k
        \right )
    \end{equation*}
    where $w_{max}$ and $w_{avg}$ are (user-specified) hyperparameters, as specified in section \ref{sec:similarity_calculation}. 
    $\vec{\tau}_{i, j}^k$ ``assumes'' the cardinality of the set of lineage vectors of the candidate tuple is $k$.
    
    \item Each long target vector $\vec{\tau}_{i, j}^{k=|V|} \in \mathbb{R}^{|A| \times |V| \times D}$ computes the desired similarity calculation via a dot product with a long candidate vector $\vec{L}_V^{n=|A|} \in \mathbb{R}^{|A| \times |V| \times D}$, under the assumption that $\vec{a}_i$ and $\vec{v}_j$ have the maximal pair-wise similarity:
    \begin{equation*}
        \vec{\tau}_{i, j}^{k=|V|} \cdot \vec{L}_V^{n=|A|} =
         \vec{\tau}_{i, j}^{|V|} \cdot \vec{L}_V^{|A|}
        = 
        % \frac{w_{max} \cdot \frac{\vec{v}_{i, j}}{\lVert \vec{v}_i \rVert} \cdot \frac{\vec{a}_{j, i}}{\lVert \vec{a}_j \rVert} + w_{avg} \cdot avg(ps)}{w_{max} + w_{avg}}
    \end{equation*}
    \begin{equation*}
        = \frac{w_{max} \cdot (\vec{\sigma}_{i, j}^{|V|} \odot \vec{L}_{A}^{|V|}) \cdot \vec{L}_V^{|A|} + \frac{w_{avg}}{|A| \times |V|} \cdot \vec{L}_{A}^{|V|} \cdot \vec{L}_V^{|A|}}{w_{max} + w_{avg}} 
        =
    \end{equation*}
    \begin{equation*}
        = 
        \frac{w_{max} \cdot \frac{\vec{a}_{i, j}}{\lVert \vec{a}_i \rVert} \cdot \frac{\vec{v}_{j, i}}{\lVert \vec{v}_j \rVert} + w_{avg} \cdot avg(ps)}{w_{max} + w_{avg}}
        =
    \end{equation*}
    \begin{equation*}
        = 
        \frac{w_{max} \cdot \frac{\vec{a}_{i}}{\lVert \vec{a}_i \rVert} \cdot \frac{\vec{v}_{j}}{\lVert \vec{v}_j \rVert} + w_{avg} \cdot avg(ps)}{w_{max} + w_{avg}}
    \end{equation*}

\end{enumerate}

\subsection{Search Method}
\begin{enumerate}
    \item Initialize $M \times M$ vector search structures $S_{1,1}, ..., S_{M,M}$, such that $S_{n, k}$ holds long \textit{candidate} vectors of dimension $n \times k \times D$ (i.e., long vectors of candidate sets of cardinality $k$, assuming target sets of cardinality $n$). Note that known approximate vector search techniques, such as LSH \cite{lsh, DBLP:journals/corr/AndoniILRS15} (or any other technique, as mentioned in section \ref{sec:intro}), can be utilized here for efficiency.
    \item For each candidate set of vectors $V$ of cardinality $|V| \in [1, M]$, and its respective construction of $M$ long \textit{candidate} vectors $\vec{L}_V^n \in \mathbb{R}^{n \times |V| \times D}$, for each $n \in [1, M]$, insert $\vec{L}_V^n$ into the vector search structure $S_{n, |V|}$.
    \item Now, suppose you are given a target set of vectors $A$ of cardinality $|A| \in [1, M]$. Construct $|A| \times k$ long target vectors $\vec{\tau}_{1, 1}^k, ..., \vec{\tau}_{|A|, k}^k$, for each $k \in [1, M]$.
    \item Hereafter, we denote a set of vectors $V$ with cardinality $k$ as $V^k$.
    \item For each $k \in [1, M]$, we look \textit{separately} for the closest (dot-product wise) candidate long vector $\vec{L}_{V_{i, j}^k}^{n=|A|} \in \mathbb{R}^{|A| \times k \times D}$ in $S_{|A|, k}$, to each $\vec{\tau}_{i, j}^k \in \mathbb{R}^{|A| \times k \times D}$, respectively. 
    Intuitively, this focuses on candidate sets of vectors of cardinality $k$.
    \item 
    Next, intuitively, we need to choose the ``best'' one among the ``winners'' of different $k$ values.
    We compute the similarity scores $\vec{\tau}_{1, 1}^k \cdot \vec{L}_{V_{1, 1}^k}^{|A|}, ..., \vec{\tau}_{|A|, 1}^k \cdot \vec{L}_{V_{|A|, 1}^k}^{|A|}, ..., \vec{\tau}_{1, k}^k \cdot \vec{L}_{V_{1, k}^k}^{|A|}, ..., \vec{\tau}_{|A|, k}^k \cdot \vec{L}_{V_{|A|, k}^k}^{|A|}$, for each $k \in [1, M]$ (a total of $\sum_{k=1}^M |A| \times k = |A| \times \sum_{k=1}^M k = |A| \times \frac{M(1 + M)}{2}$ computations). The one yielding the highest score, say $\vec{L}_{V_{{\Tilde{i}, {\Tilde{j}}}}^{\Tilde{k}}}^{|A|}$, identifies the desired candidate set of vectors $V_{{\Tilde{i}, {\Tilde{j}}}}^{\Tilde{k}}$, according to our set-oriented similarity formula. The generalization of finding the $u$ closest sets is straightforward.
\end{enumerate}

\subsection{Discussion}\label{sec:varying-cadinalities-discussion}
\begin{itemize}
    \item Note that if we keep track for which cardinalities (between 1 and $M$) there are sets of this cardinality among the ones we search, we can reduce the number of long vectors we construct and separately search for each such cardinality.
    \item In addition, the maximum cardinality $M$ need only be known in advance for the target sets (those for which nearest neighbor sets are searched) and the database of sets of vectors may admit new sets of vectors of arbitrary cardinalities.  This can be relaxed if we dynamically perform additional construction (of long candidate vectors) once a “new target set cardinality” $M' > M$ appears.
    \item At any rate, the number of relevant cardinalities (those actually appearing) of stored sets of vectors may affect the efficiency of the search method, namely, the more cardinalities there are, the more long target vectors need be constructed.
\end{itemize}
\section{Proof of Correctness}\label{sec:correctness-proof}

\subsection{Constant Cardinality of Sets of Vectors}\label{sec:proof-constant}
Let $A$ be a target set of vectors of cardinality $N$, with corresponding $N^2$ long target vectors $\vec{\tau}_{1, 1}, ..., \vec{\tau}_{N, N}$. Let us denote the set of all the candidate vector sets (of cardinality $N$) for $A$ by $C_A$.

\begin{definition}\label{def:opt-a}
    % Let $A$ be a target set of vectors of cardinality $N$, and 
    We define $OPT_A = \{V \in C_A \;|\; \operatorname{sim}(A, V)$ is maximal$\}$. That is, $OPT_A$ is a set of the \textit{optimal} candidate vector sets for $A$ with a maximal similarity to $A$. 
\end{definition}

\begin{definition}\label{def:tau-set}
    % Let $A$ be a target set of vectors of cardinality $N$, with corresponding $N^2$ long target vectors $\vec{\tau}_{1, 1}, ..., \vec{\tau}_{N, N}$. 
    For each $V \in C_A$ we define a set of long target vectors $\Tau_V = \{\vec{\tau}_{i, j} \;|\; i,j \in [1, N] \:\wedge\: \vec{a}_i \in A \:\wedge\: \vec{v}_j \in V \:\wedge\:  \frac{\vec{a}_{\Tilde{i}}}{\lVert \vec{a}_{\Tilde{i}} \rVert} \cdot \frac{\vec{v}_{\Tilde{j}}}{\lVert \vec{v}_{\Tilde{j}} \rVert} 
    % = max(\{\frac{\vec{a}_{k}}{\lVert \vec{a}_k \rVert} \cdot \frac{\vec{v}_{l}}{\lVert \vec{v}_l \rVert} \;|\; k,l \in [1, N]\}) 
    = max(ps)\}$. Here, $ps$ is the set of pair-wise cosine similarities between a pair of vectors, one taken from set $A$ and one taken from set $V$.
\end{definition}

\begin{lemma}\label{lemma1}
    % Let $A$ be a target set of vectors of cardinality $N$, with corresponding $N^2$ long target vectors $\vec{\tau}_{1, 1}, ..., \vec{\tau}_{N, N}$.
    Let $V \in C_A$ be a candidate set of vectors for $A$ with a corresponding long candidate vector $\vec{L}_V$.
    Then, $\forall i,j \in [1, N]: \operatorname{sim}(A, V) \geq \vec{\tau}_{i, j} \cdot \vec{L}_V$.
\end{lemma}

\begin{proof}
    Consider a pair of indices $i,j \in [1, N]$. By definition of set-set similarity (see formula in section \ref{sec:similarity_calculation}):
    \begin{equation*}
        \operatorname{sim}(A, V) \underset{def.}{=} \frac{w_{max} \cdot max(ps) + w_{avg} \cdot avg(ps)}{w_{max} + w_{avg}} =
    \end{equation*}
    \begin{equation*}
        \underset{ps}{=}
        \frac{w_{max} \cdot max(\{\frac{\vec{a}_{k}}{\lVert \vec{a}_k \rVert} \cdot \frac{\vec{v}_{l}}{\lVert \vec{v}_l \rVert} \;|\; k,l \in [1, N]\}) + w_{avg} \cdot avg(ps)}{w_{max} + w_{avg}} \geq
    \end{equation*}
    \begin{equation*}
        \underset{max}{\geq} 
        \frac{w_{max} \cdot \frac{\vec{a}_{i}}{\lVert \vec{a}_i \rVert} \cdot \frac{\vec{v}_{j}}{\lVert \vec{v}_j \rVert} + w_{avg} \cdot avg(ps)}{w_{max} + w_{avg}} = \vec{\tau}_{i, j} \cdot \vec{L}_V
    \end{equation*}
    where the last equality stems from the construction shown in (\ref{calc:tau-dot-product-constant}) of section \ref{sec:long_vectors_construction_constant}.
\end{proof}

\begin{lemma}\label{lemma2}
    % Let $A$ be a target set of vectors of cardinality $N$, with corresponding $N^2$ long target vectors $\vec{\tau}_{1, 1}, ..., \vec{\tau}_{N, N}$.
    Let $V \in C_A$ be a candidate set of vectors for $A$ with a corresponding long candidate vector $\vec{L}_V$.
    Then, $\forall \vec{\tau}_{i, j} \in \Tau_V: \operatorname{sim}(A, V)$ $= \vec{\tau}_{i, j} \cdot \vec{L}_V$.
\end{lemma}

\begin{proof}
    Consider a long target vector $\vec{\tau}_{i, j} \in \Tau_V$.
    % Let $a_{\Tilde{i}} \in A, v_{\Tilde{j}} \in V$ be the pair of vectors with maximal pair-wise cosine similarity, amongst all pairs of vectors, one from $A$ and one from $V$, respectively. I.e.,
    % \begin{displaymath}\label{eqn:maximal_ps}\tag{$@$}
    %     \frac{\vec{a}_{\Tilde{i}}}{\lVert \vec{a}_{\Tilde{i}} \rVert} \cdot \frac{\vec{v}_{\Tilde{j}}}{\lVert \vec{v}_{\Tilde{j}} \rVert} = max(\{\frac{\vec{a}_{k}}{\lVert \vec{a}_k \rVert} \cdot \frac{\vec{v}_{l}}{\lVert \vec{v}_l \rVert} \;|\; k,l \in [1, N]\}) = max(ps)
    % \end{displaymath}
    % Now, let $\vec{L}_V \in S$ ($S$ is the long candidate vectors search structure) be the long candidate vector that is associated with V.
    \begin{equation*}
        \Rightarrow \operatorname{sim}(A, V) \underset{def.}{=} \frac{w_{max} \cdot max(ps) + w_{avg} \cdot avg(ps)}{w_{max} + w_{avg}} =
    \end{equation*}
    \begin{equation*}%\label{eqn:similarity_equality_theorem1}\tag{$\$$}
        % \underset{\eqref{eqn:maximal_ps}}{=}
        \underset{\text{Definition \ref{def:tau-set}}}{=}
        \frac{w_{max} \cdot \frac{\vec{a}_{i}}{\lVert \vec{a}_{i} \rVert} \cdot \frac{\vec{v}_{j}}{\lVert \vec{v}_{j} \rVert} + w_{avg} \cdot avg(ps)}{w_{max} + w_{avg}} = \vec{\tau}_{i, j} \cdot \vec{L}_V
    \end{equation*}
    where the last equality stems from the construction shown in (\ref{calc:tau-dot-product-constant}) of section \ref{sec:long_vectors_construction_constant}. 
    % $\vec{\tau}_{\Tilde{i}, \Tilde{j}}$ is hereafter denoted as $\vec{\tau}_V$, for brevity.
    % \begin{equation*}\label{eqn:similarity_equality_theorem}\tag{$*$}
    %     \Rightarrow \operatorname{sim}(A, V) \underset{\eqref{eqn:similarity_equality_theorem1}}{=} \vec{\tau}_{\Tilde{i}, \Tilde{j}} \cdot \vec{L}_V = \vec{\tau}_V \cdot \vec{L}_V
    % \end{equation*}
\end{proof}

\begin{theorem}
    % Let $A$ be a target set of vectors of cardinality $N$, with corresponding $N^2$ long target vectors $\vec{\tau}_{1, 1}, ..., \vec{\tau}_{N, N}$.
    % Let $V \in OPT_A$ be a candidate set of vectors for $A$ of cardinality $N$, such that $\operatorname{sim}(A, V)$ is maximal (amongst all the candidate sets for $A$).
    The search method of section \ref{sec:search_method_constant} correctly identifies and outputs a candidate set of vectors $V' \in OPT_A$ (see Definition \ref{def:opt-a}) in the final step (\ref{search:closest-dot-products-output}).
\end{theorem}

\begin{proof}
    \par We first prove that $(*)$ there exists an optimal set $\Hat{V} \in OPT_A$ such that $\vec{L}_{\Hat{V}}$ is discovered in step (\ref{search:closest-dot-products-search}) of section \ref{sec:search_method_constant} when considering some long target vector $\vec{\tau}_{\Hat{V}} \in \Tau_{\Hat{V}}$. Then, we conclude that $(**)$ the output in the final step (\ref{search:closest-dot-products-output}) of section \ref{sec:search_method_constant} is an optimal candidate set $V' \in OPT_A$, which is associated with the long candidate vector $\vec{L}_{V'}$.
    % $\vec{\tau}_V \cdot \vec{L}_V$ yields the highest similarity score, amongst all the pairs of long target and candidate vectors, retrieved in step (\ref{search:closest-dot-products-search}) of section \ref{sec:search_method_constant}. 
    % an optimal candidate set $V \in OPT_A$, which is associated with the long candidate vector $\vec{L}_V$, is output in the final step (\ref{search:closest-dot-products-output}) of section \ref{sec:search_method_constant}, as required.
    \\
    \par $(*)$ Let us assume, by way of contradiction, that none of the optimal sets $V \in OPT_A$ (via its associated long candidate vector $\vec{L}_V$) is discovered in step (\ref{search:closest-dot-products-search}) of section \ref{sec:search_method_constant} when considering any of its respective long target vectors $\vec{\tau}_V \in \Tau_V$. I.e., for each $V \in OPT_A$ and for each $\vec{\tau}_V \in \Tau_V$ there is a long candidate vector $\vec{L}_X \in S$ such that $\vec{\tau}_V \cdot \vec{L}_X > \vec{\tau}_V \cdot \vec{L}_V$ ($S$ here is the set of vectors in the search structure).
    \begin{equation*}
        \Rightarrow \operatorname{sim}(A, X) \underset{\text{Lemma \ref{lemma1}}}{\geq} \vec{\tau}_V \cdot \vec{L}_X > \vec{\tau}_V \cdot \vec{L}_V \underset{\text{Lemma \ref{lemma2}}}{=} \operatorname{sim}(A, V)
    \end{equation*}
    \begin{equation*}
        \Rightarrow \operatorname{sim}(A, X) > \operatorname{sim}(A, V)
    \end{equation*}
    which is a contradiction to the maximality of $\operatorname{sim}(A, V)$.
    Consequently, we deduce that there exists a set $\Hat{V} \in OPT_A$ and a long target vector $\vec{\tau}_{\Hat{V}} \in \Tau_{\Hat{V}}$ such that $\vec{L}_{\Hat{V}}$ is discovered in step (\ref{search:closest-dot-products-search}) of section \ref{sec:search_method_constant}
    as the closest (dot-product wise) long candidate vector in $S$ to $\vec{\tau}_{\Hat{V}}$.
    \par $(**)$ Let us assume, by way of contradiction, that the output in the final step (\ref{search:closest-dot-products-output}) of section \ref{sec:search_method_constant} is \textit{not} an optimal set from $OPT_A$. I.e., in particular, there is a pair of indices $x, y \in [1, N]$, with respective long target and candidate vectors $\vec{\tau}_{x, y}$ and $\vec{L}_{x, y}$, such that $\vec{\tau}_{x, y} \cdot \vec{L}_{x, y} > \vec{\tau}_{\Hat{V}} \cdot \vec{L}_{\Hat{V}}$. We denote the set of vectors that is associated with $\vec{L}_{x, y}$ as $V_{x, y} \notin OPT_A$.
    \begin{equation*}
        \Rightarrow \operatorname{sim}(A, V_{x, y}) \underset{\text{Lemma \ref{lemma1}}}{\geq} \vec{\tau}_{x, y} \cdot \vec{L}_{x, y} > \vec{\tau}_{\Hat{V}} \cdot \vec{L}_{\Hat{V}} \underset{\text{Lemma \ref{lemma2}}}{=} \operatorname{sim}(A, \Hat{V})
    \end{equation*}
    \begin{equation*}
        \Rightarrow \operatorname{sim}(A, V_{x, y}) > \operatorname{sim}(A, \Hat{V})
    \end{equation*}
    which is a contradiction to the maximality of $\operatorname{sim}(A, \Hat{V})$.
    % We conclude that $\vec{\tau}_{\Hat{V}} \cdot \vec{L}_{\Hat{V}}$ yields the highest similarity score amongst the group $\{\vec{\tau}_{i,j} \cdot \vec{L}_{i,j} \;|\; i,j \in [1, N]\}$, where $\vec{L}_{i,j}$ is the closest long candidate vector in $S$, to each $\vec{\tau}_{i,j}$, respectively, as retrieved in step (\ref{search:closest-dot-products-search}) of section \ref{sec:search_method_constant}. 
    Hence, we conclude that some set $V' \in OPT_A$, which is associated with the long candidate vector $\vec{L}_{V'}$, is output in the final step (\ref{search:closest-dot-products-output}) of section \ref{sec:search_method_constant}, as required.
\end{proof}

\subsection{Sets of Vectors of Varying Cardinalities}\label{sec:proof-varying}
In section \ref{sec:proof-constant} we provide a proof for the special case, where all sets of vectors are of a constant cardinality $N$. The proof of the general case, with sets of vectors of varying cardinalities, is much more involved and requires a lot more bookkeeping; but, it is essentially the same proof as that for the special case.

\subsection{Approximate Vector Search}
We note that the proof in section \ref{sec:proof-constant} is correct under the assumption of using an exact vector search method in step (\ref{search:closest-dot-products-search}) of section \ref{sec:search_method_constant}. The correctness and accuracy with an approximate search method depend only on the correctness and accuracy of said search method.

\section{Proof of Concept Experimentation}\label{sec:experiments}

In this section, we first describe experimental results that \textit{corroborate} the correctness of the proof from section \ref{sec:correctness-proof}. Then, we show that our proposed method can be significantly accelerated by using an approximate (quantization based) vector search algorithm \cite{guo2020accelerating}, while maintaining a very high ``$k$-closest sets'' recall score. We conclude that this performance boost makes our method \textit{practical}, and \textit{scalable} to massive databases of sets of vectors.

\subsection{Experimental Setup}
\subsubsection{Dataset}
To the extent of our knowledge, there is no current benchmark to compare the performance of nearest neighbor search algorithms on datasets of sets of vectors.
As in \cite{guo2020accelerating}, we choose the \texttt{Glove1.2M} dataset for benchmarking, for its data distribution properties (low correlation across dimensions and equal variance in each dimension).
This dataset contains roughly 1.2 million \textsc{GloVe} \cite{pennington2014glove} word vectors, each of dimension $100$, trained on a twitter-derived corpus, and is a publicly available benchmark at \href{http://ann-benchmarks.com}{\texttt{ann-benchmarks.com}} \cite{AUMULLER2020101374}.
% As mentioned in \cite{guo2020accelerating}, modern large-scale databases of vectors are often created with neural network embeddings learned by minimizing some training task. This typically leads to the following nice properties: 
% \begin{enumerate*}
%     \item Low correlation across dimensions.
%     \item Equal variance in each dimension.
% \end{enumerate*}
To construct a database of sets of vectors of a constant cardinality, we choose the hyperparameter $N=3$ and divide the \texttt{Glove1.2M} dataset into roughly 400k sets of $N=3$ vectors each. The dataset has additional 10k test vectors, which we also transformed into around 3.3k test sets of $N=3$ vectors each.

\subsubsection{Metrics}
Throughout the experiments, we measure the recall, i.e., the fraction of true $k$-closest neighbors found, for different values of $k$, on average over all test queries, against the time it takes to serve a single search query, on average over all test queries. The true $k$-closest neighbors for each of the test queries are calculated with an exact brute-force search on vector sets implementation, using the vector set - vector set similarity formula from section \ref{sec:similarity_calculation}.

\subsubsection{Hardware Setup}
Our benchmarks are all conducted on an Intel Xeon 5220R machine with a single CPU thread (unless explicitly stated otherwise).

\subsection{Experiments}
\subsubsection{Exact Search}
A naive brute-force search implementation, using long vectors (as described in section \ref{sec:constant_cardinality_vector_search}) was implemented to compare the $k$-closest retrieved by this method, against the true neighbors, for different values of $k \in [1,10]$, over all test queries. The results showed perfect outputs, in terms of recall (1.0) and the order of the returned $k$-closest neighbors. This gives an experimental corroboration to the correctness of the long vectors method.

\subsubsection{Approximate Search}
\begin{figure}
    % \begin{center}
    \begin{adjustbox}{width=\columnwidth,center}
        % \inputpgf{graphics}{recall-time.pgf}
        \input{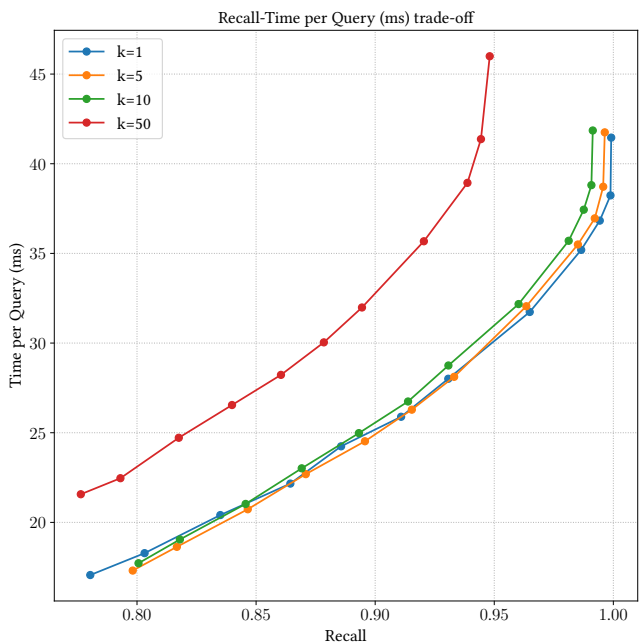}
    \end{adjustbox}
    % \end{center}
    \caption{Recall-Time per query (ms) trade-off for approximate $k$-closest task (with different $k$ values)
    % , using an approximate search algorithm on vector sets
    .}
    \label{fig:recall-time}
\end{figure}

First, for reference, we note that an exact brute-force search on vector sets takes around 2700ms per query, on average over all test queries. This brute-force search implementation involves a highly optimized (single core) similarity calculation, for a target vector set against all vector sets in the database, using efficient multidimensional array multiplications. The choice of $k$ for the $k$-closest task, with a brute-force exact search, does not greatly affect the running time since the bottleneck is in the similarity calculation and sorting phases. In Figure \ref{fig:recall-time} we present a time per query vs. recall curve (note the trade-off), using an approximate search algorithm on long vectors, for different choices of $k$. Each testing point on a curve was derived using Google's approximate (quantization based) vector search algorithm (ScaNN) \cite{guo2020accelerating}, and tweaking its \texttt{num\_leaves\_to\_search} hyperparameter. Specifically, for $k = 10$ we get a 0.991 mean recall score, with a 0.033 std, over all test queries at just 41.9ms per query. I.e., we get a speedup of about $64\scriptstyle{\times}$ using an approximate search over long vectors, while maintaining a \textit{very} high recall score. It should be noted that additional speedup can be gained with parallelization (on an 8-core  Xeon 5220R machine) and batching of the search queries. For example, choosing $k = 10$ with a mean recall of 0.991 we go down to 15.8ms per query on non-batched queries, and down to 6.2ms per query on batched queries.
\section{Conclusions}
\label{chap:conclusions}

A prevailing similarity measure between vectors is the cosine distance. For this measure, there are numerous efficient approximate search methods that given a vector (target vector) and a set of vectors, locate the closest (or $k$ closest) vectors in the set to the given target vector. However, there are applications that need to measure similarity between sets of vectors. Such a case was exhibited in \cite{DBLP:conf/vldb/LeybovichS20} that also defined a similarity measure that balances between the maximum cosine distance amongst pairs of vectors, one from each set, and the average pair-wise similarity between pairs of vectors, one from each set. This paper presents two methods for efficiently reducing the vector set - vector set similarity to vector-vector similarity. This enables using any approximate vector similarity search algorithm to, given a set of vectors, efficiently search among sets of vectors for the closest set or $k$-closest sets. This is first done assuming all considered sets have the same cardinality, and then generalizing to sets of vectors with varying cardinalities, up to a given maximum cardinality (see discussion in Section \ref{sec:varying-cadinalities-discussion}). We prove the correctness of the first of these methods; the proof can be generalized to prove the correctness of the second method. Finally, we present a series of experiments that corroborate the correctness of the long vectors method, and show its usefulness for an efficient search of vector sets in a practical setting.
% \input{main/chap4:per_column_lineage_vectors}
% \input{main/chap5:improving_lineage_embeddings}
% \input{main/chap6:experimental_evaluation}

% \input{main/conclusions}
%
% Add any appendices here; they must come _before_ the bibliography
%
% \appendix
% \noappendicestocpagenum
% \addappheadtotoc
% \include{main/appendix_efficient_vector_search}
% \include{main/appendix_bfpdb_schema}
% \include{main/appendix_movielens_schema}
% \include{main/appendix_bfpdb_mvws}

% Back Matter
% ------------

% The following command will typeset the bibliography,
% then typeset the Hebrew part of the thesis:
% - Cover page
% - Title page
% - Acknowledgements page
%  (NO table of contents or list of figures in Hebrew)
% - (Extended) abstract (1000-2000 words)
%
% based on information you've provided in the thesis-fields file
% (including the relative paths to your bib files). The Hebrew
% content will be typeset in _reverse_page_order_, i.e. first
% in the file will be the last page of the abstract, and the
% Hebrew cover page will be the last page of the file.
%
% \makebackmatter

% The resulting PDF can be printed and taken straight to binding,
% i.e. you do not need to flip any pages anywhere. Of course,
% mind the LaTeX error and warning messages, overfull hboxes etc.

%\clearpage

\bibliographystyle{ACM-Reference-Format}
\bibliography{back/general}

\end{document}